\documentclass[journal]{IEEEtran}
\usepackage{url}
\usepackage{framed}
\usepackage{hyperref}
\usepackage{cite}
\usepackage{graphicx}
\usepackage{comment}
\usepackage{booktabs}

\usepackage{ifthen}
\newboolean{showcomments}
\setboolean{showcomments}{true}

\ifCLASSINFOpdf

\begin{document}
\title{A New Definition of Demand Response in the Distributed Energy Resource Era}

 \author{Johanna L.\ Mathieu,~\IEEEmembership{Senior Member,~IEEE,} Gregor Verbi\v{c},~\IEEEmembership{Senior Member,~IEEE,} Thomas Morstyn,~\IEEEmembership{Senior Member,~IEEE,}
 Mads Almassalkhi,~\IEEEmembership{Senior Member,~IEEE,}
Kyri Baker,~\IEEEmembership{Senior Member,~IEEE,}
Julio Braslavsky,~\IEEEmembership{Senior Member,~IEEE,}
Kenneth Bruninx,~\IEEEmembership{Member,~IEEE,}
Yury Dvorkin,~\IEEEmembership{Member,~IEEE,}
Gregory S. Ledva,~\IEEEmembership{Member,~IEEE,}
Nariman Mahdavi,
Hrvoje Pand\v{z}i\'{c},~\IEEEmembership{Member,~IEEE,}
Alessandra Parisio,~\IEEEmembership{Senior Member,~IEEE,}
Vedran Peri\'{c},~\IEEEmembership{Member,~IEEE}%
\thanks{This paper is an outcome of the IEEE Power \& Energy Society (PES) Task Force on ``Demand Response in the DER Era" co-sponsored by IEEE PES Power System Planning, Operation and Economics (PSOPE) Technical Committee and the IEEE PES Smart Buildings, Loads and Customer Systems (SBLCS) Technical Committee co-chaired by G.\ Verbi\v{c} (PSOPE) and J.L.\ Mathieu (SBLCS). T.\ Morstyn is the secretary. J.L.\ Mathieu is with the University of Michigan - Ann Arbor, USA, {\tt jlmath@umich.edu}. G. Verbi\v{c} is with the University of Sydney, Australia, {\tt gregor.verbic@sydney.edu.au}.  T.\ Morstyn is with the University of Oxford, UK {\tt thomas.morstyn@ed.ac.uk}. }
}

\maketitle

\begin{abstract}
Demand response is a concept that has been around since the very first electric power systems. However, we have seen an explosion of research on demand response and demand-side technologies in the past 30 years, coinciding with the shift towards liberalized/deregulated electricity markets and efforts to decarbonize the power sector. Now we are also seeing a shift towards more distributed/decentralized electric systems; we have entered the era of ``distributed energy resources," which require new grid management, operational, and control strategies. Given this paradigm shift, we argue that the concept of demand response needs to be revisited, and more carefully/consistently defined to enable us to better utilize this massive resource for economic, technical, environmental, and societal aims. In this paper, we survey existing demand response definitions, highlight their shortcomings, propose a new definition, and describe how this new definition enables us to more effectively harness the value of demand response in modern power systems. We conclude with a demand response research agenda informed by a discussion of demand response barriers and enablers.
\end{abstract}


\IEEEpeerreviewmaketitle

\section{Introduction}
\label{sec:intro}
\IEEEPARstart{T}{he} concept of demand response (DR) has been around since the advent of electric power grids. Some of the first power systems used time-of-use (TOU) pricing. For example, in New York City in the 1890s, Samuel Insull used TOU pricing (signals) to entice new off-peak customers to maximize the load factor and increase the overall electricity demand~\cite{nyc_2014}. Perhaps the first widely-used load control program was employed in 1940s to curtail the load during peak periods for system reliability via ripple control~\cite{ripple_1948}. A decade later, in Australia, a controlled load tariff was added to shift the load of electric hot water heaters to low-demand periods and provide financial benefits to end-users~\cite{ausgrid_dm}. With such programs still in operation, DR definitions have also largely remained the same for almost 60 years, retaining a primary focus on economic motivations and system reliability during peak demand periods~\cite{ieee_dr_survey_2015, Siano_dr_survey_2014}. 

However, the power system is changing with 1) the shift to liberalized/deregulated electricity markets in the US, Europe, and Australia starting in the 1990s; 2) a focus on climate change, efficiency, and decarbonization; and 3) a paradigm shift from centralized to distributed energy resources (DERs), which include distributed generation, storage, and flexible loads, which could be customer-sited (i.e., consumer energy resources (CERs)) or grid/utility-sited. These resources enable capabilities that go beyond the traditional concept of DR. 

These changes require us to revisit the concept of DR. However, a fundamental challenge to this task is that there are enumerable competing definitions of DR. Moreover, there is a significant amount of demand-response-adjacent terminology emerging including demand (side) flexibility, demand side management (DSM), demand side resource, demand management systems, distributed energy resource management system (DERMS), virtual power plant (VPP), virtual battery, behind-the-meter (BTM) and front-of-the-meter resources, DERs, etc.

We argue that competing definitions and a ``clutter" of related concepts can undermine the value and effectiveness of DR, muddling our understanding of DR, its uses, and its benefits. Therefore, we propose a new definition to clarify DR, with an aim of defining it as broadly as possible to enable us to capture its full potential spanning economic, technical, environmental, and social objectives.

In this paper, we first survey existing DR definitions across the world and compare them, highlighting trends and gaps. We then propose our new definition that broadens the scope of DR and describe how this definition enables us to more effectively harness the benefits of DR through a discussion of DR services and value streams, mechanism design, and enabling technologies. Finally, we provide a discussion of the technical, regulatory, and social  barriers and enablers of DR, and we use this as context to propose a research agenda for DR, identifying critical topics -- academic and applied -- that should be addressed to effectively advance DR.

\section{Defining Demand Response}
\label{sec:survey}

\subsection{Comparison of existing definitions} 
\label{DEF}

There are many competing definitions of DR. For example, Wikipedia defines DR as ``\textit{a change in the power consumption of an electric utility customer to better match the demand for power with the supply}"~\cite{wiki_2024}. This definition is based on a BBC News article's definition~\cite{bbc_2023}. In contrast, the International Energy Agency (IEA) website defines DR as, ``\textit{balancing the demand on power grids by encouraging customers to shift electricity demand to times when electricity is more plentiful or other demand is lower, typically through prices or monetary incentives}"~\cite{ieadef_2024}. Different definitions and levels of detail in definitions lead us to different understandings of the term and different notions of value that DR brings.

One way in which we can compare definitions is in terms of the way DR is motivated. We analyzed 16 DR definitions across the US, Europe, and Australia. The definitions are provided in~\cite{appen} and Table~\ref{tab:Kyri_Table} compares them in terms of the motivations for DR (grid reliability, economic, and/or environmental). We recognize that interpretations of each definition and categorization may be subjective. Our process involved analyzing the ``concise'' definition from each source (e.g., typically the sentence or sentences after a statement of the form ``Demand response is a...'') rather than including the ``extended'' definition (e.g., all of the content within the referenced document or webpage). ``Grid Reliability'' was indicated if the definition mentioned ``reliability'' or related concepts (e.g., ``helping to balance the energy system'').  Similarly, ``Economic'' and ``Environmental'' motivations were indicated for definitions that mention goals related to cost savings (including terms such as ``cost-effective'') and emissions/environmental goals, respectively. In the table, we delineate between definitions that explicitly mention these motivations versus definitions that imply these motivations.

Most definitions motivate DR based on its ability to improve grid reliability and economics, i.e., its ability to reduce wholesale electricity prices. Only recently, there have been some attempts to expand the DR definition to include environmental motivations. Our proposed definition, which will be presented in Section~\ref{sec:new-dr-definition}, does not explicitly or implicitly mention the motivations for DR -- we aimed to develop a definition that simply defines what DR \emph{is} rather than what it \emph{aims to do}.  We will discuss our reasons for this in Section~\ref{sec:new-dr-definition}.

\begin{table*}
\begin{center}
\caption{Demand response motivations across a variety of definitions. x = explicit, (x) = implicit.}
\vspace{-.3cm}
\label{tab:Kyri_Table} 
\begin{tabular}{lccc}
\toprule
Source & Grid Reliability & Economic & Environmental\\
\midrule
US DOE (2006) \cite{DOE_Def06}                      & x     & x     & \\ 
FERC Orders 745 (2011) \cite{FERC745} \&  2222 (2020) \cite{FERC2222}             &       & x     & \\ 
CPUC (2017) \cite{CPUC_Def17}                       & x     & x     & \\ 
EU Directive 944 (2019)  \cite{eu_directive}        &       & x     & \\ 
UK OFGEM (2021)  \cite{OFGEM}                       & (x)   & x     & x \\ 
Belgium (2022)* \cite{belgium_dr}                    & x     & x     & x \\ 
IEA (2024) \cite{ieadef_2024}                       & x     & x     &  \\ 
National Grid ESO UK (2024) \cite{ESO_Def24}        & (x)   & x     & x \\ 
National Grid USA (2024) \cite{NatGridUS_Def24}     &       & x     & x \\ 
AEMC (2024)  \cite{AEMC_Def24}                      &       &       & \\ 
ARENA (2024) \cite{ARENA_Def24}                     & x     &       &  \\ 
AEMO (2024) \cite{AEMO_Def24}                       &       &       & \\ 
IESO (2024) \cite{IESO_Def24}                       &       &       & \\ 
NYISO (2024) \cite{NYISO_Def24}                     &       &       & \\ 
PJM (2024) \cite{PJM_Def24}                         & x     & x     & \\
\midrule
Proposed Definition & & &  \\ 
\bottomrule
\end{tabular}
\\ \vspace{.2cm}
*Definition not available in English; assessment based on English translation provided in~\cite{appen}.
\vspace{-.3cm}
\end{center}
\end{table*}

We also compare definitions in terms of how DR actions are defined. Table~\ref{tab:Kyri_Table2} compares the same 16 definitions in terms of six specifications around DR actions, specifically, whether the actions are {\bf 1) not limited to load reduction}, but could also include load shifting, load increasing, and/or net production (from, e.g., customer-sited generation, storage, and/or load); {\bf 2) specified as voluntary}, meaning DR participants may choose whether or not to participate in DR programs and/or choose whether or not to take DR actions (i.e., customers are not forced to adjust their load); {\bf 3) specified as active}, meaning DR actions are active changes to consumption, not passive responses to changes in grid states (e.g., conservation voltage reduction or inherent inertial response provide a ``passive'' response); and {\bf 4) specified as temporary}, meaning the action is not sustained indefinitely like long-term energy efficiency upgrades. We also assess whether or not the DR definition {\bf 5) specifies the location of the responding asset} (e.g., customer-sited or behind the meter, as opposed to a network asset) and whether {\bf 6) the DR signal is broadly defined}, meaning the definition {\em both} mentions the need for a DR ``signal" and defines it broadly. Commonly DR signals are external functional signals (on/off, setpoint adjustment) or price/incentive signals, but could also be interpreted to include measurements, e.g., of frequency and voltage. We argue that all six of these specifications are critical for defining DR. 

In most definitions, DR is not limited to load reduction, but several North American definitions specify DR actions as only load reductions. Most definitions explicitly or implicitly require DR actions to be voluntary and temporary; however, these specifications are missing from a number of definitions. Only one definition explicitly states that DR should be an active response, though this is implied in nearly all definitions, e.g., by mentioning changes taken by customers. Similarly, no definitions explicitly specify the location of the responding assets, but nearly all imply their location, e.g., by mentioning the customer. Few definitions both mention the need for a DR signal and define that signal broadly. When it is mentioned, but not defined broadly, it is often specified as a price or incentive.

\begin{table*}
\caption{Demand response action specifications across a variety of definitions. x = explicit, (x) = implicit.}
\vspace{-.3cm}
\label{tab:Kyri_Table2} 
\centering
\begin{tabular}{lcccccc}
\toprule
        & Not Limited to & Specifies & Specifies        & Specifies & Specifies Location of & DR Signal\\
Source  & Load Reduction & Voluntary & Active Response  & Temporary & Responding Assets     & \hspace{-.3cm}Broadly Defined \\
\midrule
US DOE (2006) \cite{DOE_Def06}                &  x  & x & (x) & (x) & (x) &  \\ 
FERC Orders 745 (2011) \cite{FERC745} \&  2222 (2020) \cite{FERC2222}   \hspace{-1cm}        &     & (x) & (x) &     & (x) &  \\ 
CPUC (2017) \cite{CPUC_Def17}                   &  x  &     & (x) &     & (x) & (x) \\ 
EU Directive 944 (2019) \cite{eu_directive}     &  x  & (x) & (x) &     & (x) & \\ 
UK OFGEM (2021) \cite{OFGEM}                    &  x  & (x) & (x) & (x) & (x) & \\ 
Belgium (2022)* \cite{belgium_dr}                &  x  &     &     &     &     & \\ 
IEA (2024) \cite{ieadef_2024}                   &  x  & (x) & (x) & (x) & (x) & \\ 
National Grid ESO UK (2024) \cite{ESO_Def24}    &  x  &     & (x) &     & (x) & x  \\ 
National Grid USA (2024) \cite{NatGridUS_Def24} &     & (x) & (x) &  x  & (x) & \\ 
AEMC (2024)  \cite{AEMC_Def24}                  & (x) &  x  &  x  &  x  & (x) & \\ 
ARENA (2024) \cite{ARENA_Def24}                 &  x  &  x  & (x) &     & (x) & \\ 
AEMO (2024) \cite{AEMO_Def24}                   &  x  &     & (x) &     &     & x \\  
IESO (2024) \cite{IESO_Def24}                   &     &     & (x) &  x  & (x) &  \\ 
NYISO (2024) \cite{NYISO_Def24}                 &     &     & (x) &  &     & \\  
PJM (2024)  \cite{PJM_Def24}                    &     &  x  & (x) &  x  & (x) & \\ 
\midrule
Proposed Definition & x & x & x & x & x & x\\ 
\bottomrule
\end{tabular}
\\ \vspace{.2cm}
*Definition not available in English; assessment based on English translation provided in~\cite{appen}.
\vspace{-.3cm}
\end{table*}

\subsection{The problem with the existing landscape of definitions}

There is generally a consensus that DR refers to a change in electricity use from normal consumption patterns by consumers, which is often made in exchange for a financial incentive. Besides this agreement, existing definitions generally differ, and sometimes they conflict. We next argue that there are a number of problems with the existing landscape of definitions, requiring a new definition. 

First, many definitions, especially those in North America, reflect notions of DR from the past, when it was used primarily for peak load reduction to improve power system reliability and reduce costs (e.g., FERC's definition: ``\textit{Demand response is a reduction in the consumption of electric energy by customers from their expected consumption in response to an increase in the price of electric energy or to incentive payments designed to induce lower consumption of electric energy}''~\cite{FERC745}). However, DR 1) can be used to {\em shape} demand; 2) can leverage not only load flexibility, but also BTM energy storage and generation; and 3) can provide other benefits such as reducing the environmental impact of electricity production/use. Limiting DR to just peak load reduction and to just motivations around reliability and/or costs undermines its full value to provide a variety of different grid services and to address a variety of different motivations.  

Second, while some definitions overly restrict DR to certain actions, signals, and/or motivations, others are too broad, which could lead to confusion over what should and should not count as DR. Some definitions are simultaneously too restrictive and too broad (e.g., NYISO's definition: ``\textit{Demand response is the act of reducing energy consumption from the grid at the direction of the NYISO}"~\cite{NYISO_Def24}, which theoretically includes utility-initiated brownouts or energy efficiency measures encouraged by NYISO, which we would argue are not DR.) To be useful, a definition should balance being broad enough to encompass all DR actions and motivations while being specific enough to rule out actions that are not DR, e.g., actions that are not voluntary, active, and temporary. 

Third, some definitions are imprecise, only implicitly specifying the key characteristics of DR. For example, UK OFGEM's definition (which uses the term ``demand side response" instead of ``demand response")  includes all DR action specifications in~Table~\ref{tab:Kyri_Table} except the broadly-defined DR signal: ``\textit{Smart meters, technologies, tariffs and services will enable consumers to change their consumption patterns to match times of cheap and abundant low carbon electricity, give consumers greater control over their energy use and comfort levels and save money by helping to balance the energy system. This is known as demand side response (DSR)}"~\cite{OFGEM}. However, most of these specifications are implicit. The definition does not limit changes to load reductions, but it only implies DR actions are voluntary, active, and temporary, and it only implies the location of the assets by mentioning the customer. While most existing definitions are centered around the utility and network needs, this definition shifts the focus to consumers and their comfort needs, and broadens the motivation for DR to also include environmental considerations. However, the definition's lack of precision, again, can lead to confusion over what should and should not count as DR.

Fourth, there are too many competing definitions of DR, which muddles our thinking about DR, its uses, and its value. Likely because it is easy to pick apart existing definitions of DR (as we have done above), it has been appealing for countries, regions, and grid operators to redefine DR to fit their needs and priorities. However, this makes it difficult to quickly come to a common understanding when discussing DR. For example, when mentioning DR in a paper, it is often necessary for the authors to provide the definition of DR they are assuming so that the readers understand how the authors see DR. It is then common for the authors and reviewers to debate that definition and whether what is being proposed is, in fact, DR. While we recognize that providing yet another definition of DR may be counter-productive, we believe that the power systems community can see the value of a common definition that is used consistently.

Fifth, because of all of the above-mentioned problems, new DR-adjacent terms have emerged to fill in gaps, which further muddles the landscape. For example, with the current focus on ``flexible technologies" and ``flexibility services", the terms demand flexibility, demand-side flexibility, and flexible demand have become common, though we would argue these simply re-brand DR. Demand-side flexibility (DSF), as defined by Smart Energy Europe (smartEn), is “\textit{the capability of any active customer to react to external signals and adjust their energy generation and consumption in a dynamic time-dependent way, individually as well as through aggregation}”~\cite{DNV}. Like most DR definitions, this definition of DSF incorporates demand-side resources responding to external signals. However, it is also broader since it explicitly mentions customer-sited generation. It also frames DR not in terms of an ``action" but in terms of a ``capability."  There are other conflicting definitions of DSF. In particular, the International Renewable Energy Agency (IRENA) defines DSF as ``\textit{the portion of demand in the system (including via electrified heat and transport) that can be reduced, increased or shifted within a specific duration}" \cite{IRENA_DSF}. In this case, DSF refers to latent potential flexibility (measured in terms of power) that could be activated through DR. This is sometimes also referred to as the ``demand responsive load" or the ``DR resource."

In summary, there is a strong need for a DR definition that is 1) broad enough to encapsulate the demand-side's full capabilities and value in responding to grid needs, 2) narrow enough to rule out actions that are not DR, 3) precise and concise in language, 4) broadly accepted by the power systems community as {\em the} definition of DR, and 5) clear in how it relates to (or should supersede) DR-adjacent concepts. Next, we propose a definition that addresses all of these gaps.

\subsection{Proposed new demand response definition and justification}
\label{sec:new-dr-definition}

\begin{framed}
  \begin{quote}
    \raggedright
  \textbf{Proposed Definition (Demand Response)}: Demand response is the actions of customer-sited energy resources downstream of metering points to \emph{voluntarily, actively, and temporarily} adjust their electricity production and/or consumption in response to signals (e.g., commands, prices, measurements).
\end{quote}  
\end{framed}

This definition is compared to the other definitions in the last rows of Tables~\ref{tab:Kyri_Table} and \ref{tab:Kyri_Table2}. Note that the definition does not specify any motivations in Table~\ref{tab:Kyri_Table}, and it includes all specifications in Table~\ref{tab:Kyri_Table2}. We have developed a definition that simply defines what DR \emph{is} rather than what it \emph{aims to do}. Specifically, this new definition avoids representing specific/narrow regulatory and operational scenarios or grid-support services. Also, being sufficiently broad (i.e., not specifying motivations) enables the definition to encompass future developments, which is critical, given the rapidly evolving related context and technology. However, the fundamental underlying actions that constitute DR should be precisely and concisely defined, and consistent.  Since the definition uses the term ``customer-site energy resources" we will henceforth refer to DR resources as CERs.

The definition is meant to be broad and inclusive of many diverse forms of DR but also rule out actions that are not DR. For example, actions that fit the proposed definition include 1)~Traditional load shedding/shifting in response to DR signals or prices (TOU, peak, and dynamic pricing);
2)~Voluntary participation in direct load control and interruptible load management programs;
3)~CERs (loads, storage, and/or distributed generation such as solar PV) providing temporal price arbitrage or participating directly in capacity, energy, or ancillary service markets;
4)~CERs providing spatial arbitrage, e.g., managing data center load across multiple sites;
5)~Single loads providing grid balancing services or contingency services;
6)~Aggregations of loads providing grid balancing or services, contingency services, or some forms of congestion management;
7)~CERs using local measurements of frequency to provide primary frequency response/droop control;
8)~CERs using local measurements of voltage to provide voltage support services;
9)~Buildings as hubs of responsive loads;
10)~Single- or bi-directional coordinated charging of electric vehicles (EVs) or electric railway systems; 
and
11)~Non-firm/flexible connection agreements (assuming the customer has voluntarily agreed) in which a customer’s consumption and/or production limits are managed via a contractual agreement~\cite{ENA_FlexibilityConnections_2021}.
We note that this list is non-exhaustive.

Actions that do {\emph not} fit the proposed definition include 1)~Energy efficiency, also known in some jurisdictions as electricity waste reduction (EWR), because actions are not meant to be temporary;
2)~Conservation Voltage Reduction (CVR), specifically, actions taken by the utility to reduce network voltages to reduce power/energy consumption because participation by customers is not voluntary or active;
3)~Under-frequency load shedding (UFLS) by the utility or system operator because participation by customers is not voluntary or active;
4)~Rotating outages/brownouts, i.e., load/generation shedding, initiated by the utility or system operator because participation by customers is not voluntary or active;
5)~Inherent inertial response (e.g., from customer-sited motors) because customer participation is not voluntary or active but rather (passively) dictated by the physics of the devices; and
6)~Actions by network-sited assets like utility-owned energy storage and voltage regulation equipment because the resource is not customer-sited. Again, this list is non-exhaustive.

\subsection{Nuances}
\label{sec:nuances--controversy}

Of course, we acknowledge that there is no perfect definition. We also acknowledge that there will be controversy about what actions fit the proposed definition and whether or not they should. Rather than attempting to clarify every type of action, we instead discuss some of these controversies here. We note that we do not have to come to a universal agreement on every action for the definition to be useful in helping us 1) understand DR and 2) increase its value to the power system.

One key point of contention may be whether or not customer-sited solar PV response (e.g., curtailment, volt/var control) should be considered DR, especially when its actions are uncoordinated with customer-sited load or energy storage. Per our definition, these actions count as DR because customer-sited resources take them temporarily, actively, and voluntarily. We believe this is consistent with how utilities generally treat BTM production as (negative) demand. However, we observe that volt-var and volt-watt automatic response modes are now mandated for new inverter-connected energy systems in some jurisdictions (e.g., since 2021 in Australia and New Zealand \cite{(aemo)23:_compliance4777,ASNZS4777.2:2020}), which makes them \emph{required} built-in responses, not voluntary, and consequently, not DR. Since these responses are automatically activated by the local voltage, they are generally temporary.

Actions taken by customer-sited fossil-fuel-burning backup generators are another point of contention. Per our definition, these actions should be considered DR. Some US states allow backup generators to participate in remunerated DR programs, but others do not because of the associated environmental impacts. It is not clear whether or not these actions should be encouraged, given the broader goals of DR, but we do not aim to settle this debate here.

\section{Value of new definition} \label{sec:valueNewDef}

We next highlight how the new definition enables us to harness the value of DR more effectively by discussing DR services and their value streams, explaining how they can be monetized via specific architectures and implementation mechanisms, and describing required technologies.

\subsection{DR Services and Value Streams}
CERs can provide 1) traditional DR services, 2) power system services through participation in bulk energy, capacity, and ancillary service markets, 3) local network services, and 4) BTM services. Existing DR definitions primarily align with 1), though some definitions admit elements of 2)-4). Our definition comprehensively covers all four service types. 

\subsubsection{Traditional services}
Historically, DR programs have been used to deal with discrete 
events, e.g., contingencies and predictable peak loads periods. Many early DR programs relied on large industrial customers that could curtail electricity-intensive processes via, e.g., interruptible load management programs~\cite{albadi_summary_2008}. Later programs incorporated large commercial loads, and leveraged TOU pricing, dynamic pricing, and demand/capacity bidding, again with a focus on peak load curtailment. In some parts of the world, residential customers participate in DR via direct load control and TOU pricing. 

\subsubsection{System services}
CERs are capable of participating alongside supply resources in energy, capacity, and ancillary service markets, though many regions do not yet fully allow this. For example, in Australia, only industrial and large commercial loads can participate in the wholesale DR mechanism~\cite{AEMC2020}.
In the US, FERC Order 2222~\cite{FERC2222} facilitates electricity market participation by DERs by requiring the Independent System Operators (ISOs) it regulates to open up their markets to DERs, but the roll-out of ISO mechanisms to support this is ongoing. Increasing renewable integration requires a more responsive demand-side resource to provide grid balancing services. While some types of industrial and large commercial loads are relatively easy to engage in peak load shedding, their individual size or end-use processes are usually not well-suited for system services. In contrast, small, continuously-responsive CERs are highly granular and, therefore, well-suited to providing balancing services when aggregated~\cite{callaway_achieving_2011}. They also present the largest untapped pool of DR resources.

\subsubsection{Local network services}
CERs can also be used to manage power flows in distribution networks by providing local network services, such as congestion management and voltage control. In contrast to bulk capacity and balancing services, which can be sourced over a large geographical area, voltage control and congestion management services in distribution networks are location-specific.
The buyer of local network services is generally the distribution system operator (DSO)/utility (also referred to as a network service provider), so network services provided by CERs can be used as a non-network or non-wires alternative to network upgrades~\cite{chew2018non}. Field demonstrations of CERs providing network services are in the early stages, with trials in Australia exploring: 1) local services  distributed energy markets, 2) bilateral procurement of network services, 3) dynamic network pricing, and 4) real-time procurement of local network support~\cite{ARENA_DEIP_2022}. In these trials, dynamic operating envelopes~\cite{liu2021grid,nazirGridAware2021} are used to communicate local network hosting capacity to DERs.   

\subsubsection{Behind-the-meter services}
BTM services aim to reduce a consumer's electricity bill or manage a local (i.e., building-level) constraint. These services can complement traditional and/or system services as CERs operated to reduce costs also perform services in response to a (price) signal designed to benefit the system. They can also complement local network services by managing local power flows. One BTM service is retail price arbitrage, where the net demand is reduced during the high-price periods and increased during the low-price periods. Retailers/utilities encourage retail price arbitrage to mitigate their price exposure in the wholesale market. A second service is peak load reduction to reduce peak load network charges (also known as demand charges or capacity tariffs), i.e., charges for the maximum average $x$-minute load over a fixed time frame. DSOs/utilities assess these charges to reflect network utilization and infrastructure costs. Managing the peak load is also important when connecting a new building/facility, as connection costs are a function of the connection power. Thus, peak load management has the triple effect of reducing energy costs, reducing peak load network charges, and reducing one-time connection costs. A third service is local constraint management, in which CERs are used to manage a physical connection constraint at a building/facility, e.g., electrical panel capacity. Often electrical codes require electrical connections to handle all feasible flows from all installed devices, and so a consumer is unable to install a new device if their system is not sized to accommodate it. However, DR could be used to manage local constraints deferring or eliminating the need to upgrade the connection.

\subsection{Mechanism design for demand response}
\label{sec:mechanism_design}

Mechanism design concerns the design of rules and institutions to achieve desired outcomes in economic and social systems~\cite{Paccagnan_Chandan_Marden_2022}. A range of DR mechanism designs have been proposed, varying by engaged participants, types of services, technologies, computation/communication requirements, level of centralization, and whether coordination is via direct or indirect negotiation~\cite{Abedrabboh_Al-Fagih_2023}. In this section, we detail DR architectures and implementation approaches. Our DR definition covers all approaches in an implementation-agnostic manner.

\subsubsection{Architectures}
\label{sec:architectures}
\begin{figure}[t]
\includegraphics[width=8cm]{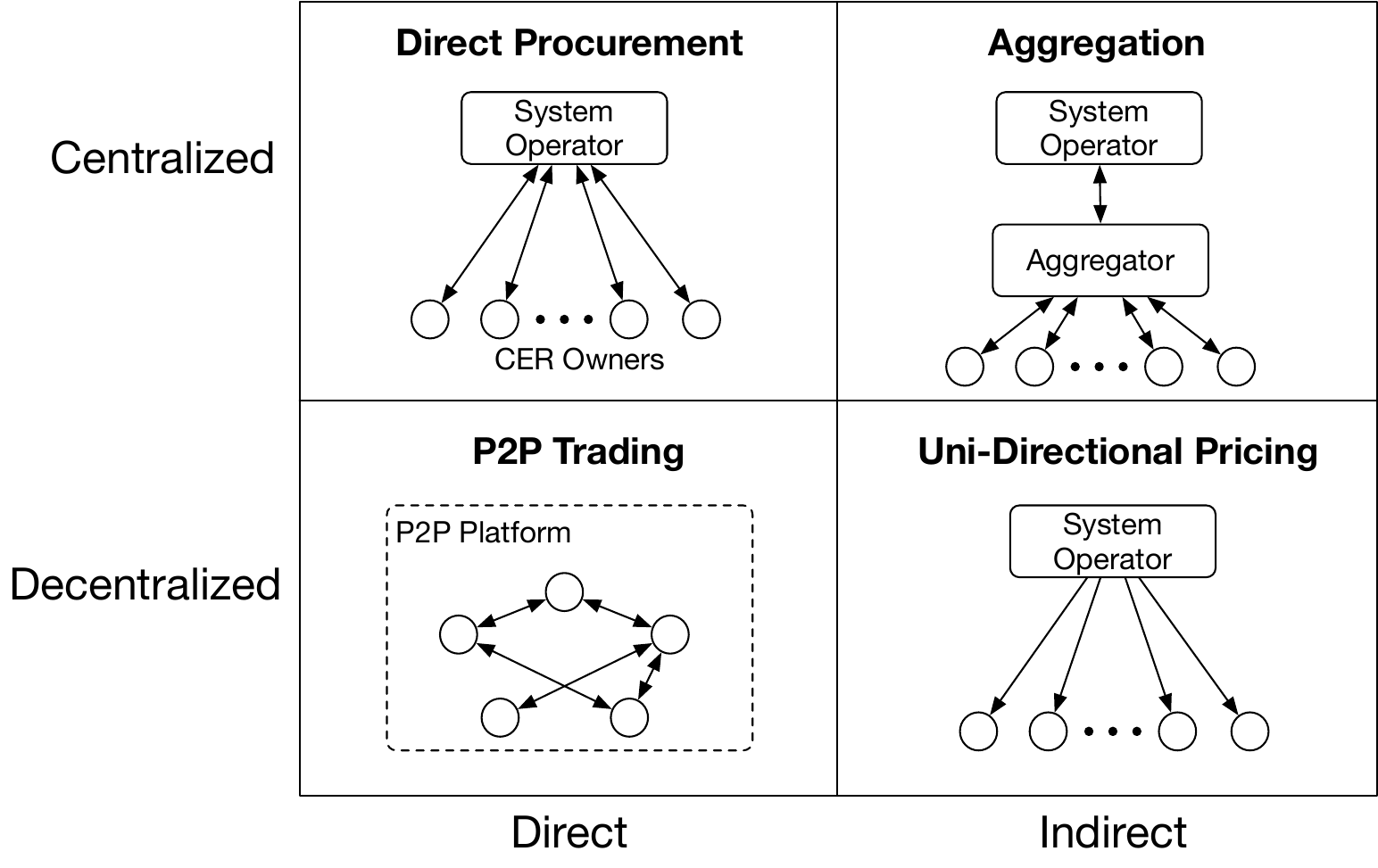}
\vspace{-.3cm}
\caption{Dominant architectures for DR mechanism design and how they vary in terms of centralization and whether coordination is direct or indirect. Specific DR strategies may blend elements of these architectures and/or include multiple entities, e.g., aggregators or P2P platforms.
}
\label{fig:MD_Archs}
\vspace{-.3cm}
\end{figure}

There are four dominant architectures for DR mechanism design (see Fig.~\ref{fig:MD_Archs}):  (i) direct procurement, (ii) aggregation, (iii) peer-to-peer (P2P) trading, and (iv) uni-directional pricing. Existing DR definitions favor uni-directional pricing and, to some extent, direct procurement, while rarely covering the other architectures.  We categorize each architecture in terms of centralization and whether coordination is direct or indirect. Centralized approaches are more inline with how power systems are currently operated; however, they usually require a high degree of information sharing leading to privacy and single-point-of-failure risks. Decentralized approaches mitigate these risks, but may reduce DR performance (e.g.,  resulting in price/power oscillations~\cite{callaway_achieving_2011}). 

    {\bf Direct procurement}: DR services can be directly procured from CERs, either by updating existing electricity market mechanisms or by developing new DR-focused mechanisms. Methods for incorporating large sources of transmission-connected DR, such as industrial sites, into markets (e.g., for frequency response and regulation) are well established~\cite{Golmohamadi_2022}.  Procuring DR services from CERs embedded within local distribution networks is more challenging as CERs are generally too small for direct procurement within regional-scale electricity markets. To address this, new centralized local energy and flexibility market designs have been proposed~\cite{Charbonnier_Morstyn_McCulloch_2022}. These local markets can enable economically efficient energy dispatch and flexibility services, but they represent a significant change to traditional retail electricity arrangements~\cite{Sahriatzadeh_Nirbhavane_Srivastava_2012} and may have significant distributional effects. Key design considerations include allocative efficiency, distributional fairness, constraint management, uncertainty handling, data privacy, computational scalability, and distribution--transmission coordination.

    {\bf Aggregation}: Directly procuring DR services from small-scale CERs requires coordinating thousands to millions of devices. To enable scalability,  CERs can be aggregated by an intermediary (aggregator or VPP) and then managed as a single dispatchable unit within electricity markets~\cite{Dielmann_Van_2003}.
    The aggregation's capabilities depend upon the aggregation's composition and coordination architecture, as well as the spatial distribution of CERs~\cite{almassalkhi2023:current}. Mechanisms are needed to share costs/profits with participating CER owners~\cite{Rahmani-Dabbagh_Sheikh-El-Eslami_2016}.    

    {\bf P2P trading}: P2P trading can be used as an alternative to centralized market mechanisms~\cite{Parag_Sovacool_2016}. Under P2P trading, CER owners and other local market participants can directly negotiate with one another. P2P transactions can involve direct energy trading or other services, such as spare export capacity~\cite{Bjarghov_Askeland_Backe_2020}. P2P trading offers advantages in terms of scalability, privacy, and CER owner autonomy~\cite{Morstyn_Teytelboym_Mcculloch_2019}. P2P trading can also support bottom-up federation to enable participation in  markets as a VPP, similar to the aggregation architecture~\cite{Morstyn_Farrell_Darby_McCulloch_2018}. Network constraint management is challenging under P2P trading since the feasibility of flows on the local network depend on the collective operation of embedded devices~\cite{Morstyn_Savelli_Hepburn_2021, 8938817}. Network constraint handling approaches include the use of P2P transaction fees~\cite{Morstyn_Teytelboym_Hepburn_McCulloch_2020, 8938817}, centralized transaction checking/approval~\cite{Guerrero_Chapman_Verbic_2019, 8938817}, integration with local flexibility markets \cite{Khorasany_Shokri}, and operating envelopes~\cite{azimDynamicOperatingEnvelopeenabled2023}. 
        
    {\bf Uni-directional pricing}: Sending uni-directional prices eliminates the need for bidirectional communication. 
    For example, TOU  tariffs  already  incentivize load shifting~\cite{Yan_Ozturk_Hu_Song_2018} and enhancing their spatio-temporal granularity can facilitate further coordination~\cite{Widergren_Marinovici_Berliner_Graves_2012}. However, the effectiveness of uni-directional pricing depends on the accuracy of demand and demand flexibility forecasts and could exacerbate demand synchronization when networks have significant amounts of embedded flexibility~\cite{Li_Wu_Oren_2014}. Nonlinear and randomized pricing arrangements have been proposed to help mitigate this~\cite{Papadaskalopoulos_Strbac_2016}.

\subsubsection{Implementation Approaches}
The architectures in Fig.~\ref{fig:MD_Archs} could be implemented through regulatory or competitive mechanisms, or a combination of both. The implementation strategy affects resource allocation, participation incentives, coordination among stakeholders, and performance.  Note that henceforth, for brevity, we often refer to the entity managing DR (i.e., the ISO, DSO, utility, retailer, aggregator, or P2P platform) as the ``DR operator."

\textbf{Regulatory mechanisms}: Regulatory mechanisms are often present in uni-directional pricing and non-market-based direct procurement. They may lead to inefficient signals and thus undermine the value of the program, which is already asymmetrically low to participants relative to DR operators. For example, TOU and dynamic pricing offer financial rewards to encourage participation, but can also cause negative effects such as cross-subsidization, failure to engage specific participant groups, and inefficiencies due to their typically coarse temporal granularity and lack of spatial granularity. Ultimately, the performance of any regulatory mechanism is based on ensuring fair and transparent operation and facilitating participant trust. However, improper design can lead to the poor outcomes and an exodus of participants.  

\textbf{Competitive mechanisms}: As opposed to regulatory mechanisms, which cannot be customized for every operational eventuality and the individual needs of each participant, competitive mechanisms provide participant-specific signals. For example, auctions enable DR procurement through competitive bidding, ensuring efficient resource allocation, price discovery, and compensation for participants. This can be achieved through participation in electricity markets via direct procurement, aggregation, or P2P trading. The economic gains of competitive mechanisms relative to regulatory mechanisms vary for different auction choices (e.g., in connection to their ability to sustain efficiency, cost recovery, revenue adequacy, or truthfulness of bidding). Additionally, competitive mechanisms facilitate communication, data exchange, and coordination, streamlining the enrollment, monitoring, and implementation of DR programs. Furthermore, competitive mechanisms naturally accommodate the diversity of choices by DR participants, particularly those related to data exchange (thus facilitating privacy) and opt-in/out options. 

\subsection{Enabling technologies} \label{sec:enablingTechnologies}

Different technologies support different DR services, architectures, and implementation approaches. We next describe the technologies that are required to enable the value streams and mechanisms consistent with our broad DR definition.

\subsubsection{Local Sensing, Actuation, and Computation} 
All DR architectures require CERs to 1) receive information via an external or locally-sensed signal and 2) be able to act (i.e., change their consumption or production). CERs may integrate additional sensors (e.g., temperature) to enable appropriate responses, data storage, and local computing needed to implement control, estimate parameters, or otherwise process data. For some DR services/architectures,  communications between CERs and advanced metering infrastructure (AMI), i.e., smart meters, which provide sensing and computing capabilities, may be useful. These functions could also be embedded within the CER itself (e.g., smart thermostat or EV charger) or within a building energy management system (EMS)~\cite{ORNL_HEMS_2017}. 

\subsubsection{Communication} 
DR typically requires communication, except when based solely on local measurements. Some of the first DR programs used ripple control, a type of power line communication, to control individual appliances directly~\cite{ripple_1948}. Early DR programs also relied on phone calls, radio announcements, or internet posts announcing DR events. Today, application programming interfaces (APIs), which define how to interact with devices/systems and provide consistent meaning to data being sent/received, are becoming common. Communication can also be possible via AMI; however, smart meters are not typically not configured for real-time communications. Also, a P2P trader or an aggregator may not have direct access to AMI, since it is typically owned by the retailer/utility.  With increasing penetration of CERs, standarized DR communication protocols are becoming more important. Examples include Open Automated Demand Response (OpenADR)~\cite{Samad_OpenADR_2016}, which facilities exchange of DR information between utilities and consumers, and IEEE 2030~\cite{basso_ieee_2014}, which facilitates device-to-device communication and integration across different systems encompassing multiple aspects of grid interoperability including communication protocols, data models, and security mechanisms for AMI, inverters, EMS, EVs, etc.

\subsubsection{Modeling and Forecasting}
Modeling and forecasting can be used by DR operators to determine the need for DR. Some DR operators simply use weather forecasts to call DR events~\cite{sce_rtp}. Others use demand forecasting, i.e., predicting future electric demand based on historical data, weather, time of day, season, etc. in lieu of demand bids to determine capacity requirements for system services. Demand forecasts can also be used together with network models to forecast, e.g., network congestion and the need for local network services. Additionally, models can be used by participants to determine demand forecasts and flexibility forecasts, and to design optimal CER responses. 

\subsubsection{CER Orchestration} \label{sec:orchestration}
CER orchestration, carried out by the DR operator, coordinates CERs to achieve specific objectives or outcomes. The concept of orchestration is closely linked to aggregation. While orchestration achieves coordination, aggregation achieves consolidation of individual CERs into a single dispatchable unit. Direct procurement and uni-directional pricing orchestration is enabled through communication protocols. P2P trading uses digital platforms that connect individual participants allowing the platform to offer response from any/all platform participants, thus effectively acting as an aggregator. For aggregation architectures, the orchestration problem could be cast as a large-scale model-based control problem and solved centrally; however, this can be computationally expensive when the aggregation includes a large number of CERs. Alternative approaches leverage distributed optimization~\cite{Molzahn_DistributedSurvey_2017} or reduced-form system models that capture essential aggregate CER dynamics~\cite{mathieu_state_2013}.

\section{Barriers and enablers}
\label{sec:barriers}

In this section, we describe DR barriers and enablers from 1) technical; 2) regulatory, policy, and market design, and 3) social perspectives, and we use this discussion to inform a DR research agenda. 

\subsection{Technical Considerations} \label{sec:TechBarriers}

Emerging technology trends may enhance DR uptake, increase the scale of existing DR implementations/systems, or increase DR functionality, better aligning DR capabilities with our broad definition of DR. Here, we revisit each category listed in Section~\ref{sec:enablingTechnologies} and discuss technology enablers that will help us more effectively harness the full value of DR. We also discuss associated limitations/barriers.

\subsubsection{Local Sensing, Actuation, and Computation} 
Grid-edge computing is an emerging technology that enables faster sensing and more substantial computing capabilities at a meter, building gateway, or CER~\cite{nationalGrid2022}. Cloud computing technology allows the data processing and storage capabilities of CERs to be augmented with cloud infrastructure~\cite{cecCloud2021}. New sensing technologies allow faster sampling of data streams and the measurement or estimation of additional quantities (e.g., voltage, current, and frequency waveforms). Grid-edge computing allows advanced algorithms to run at the meter, building gateway, or CER, enabling more complex DR functionality, novel control architectures, and the potential for more targeted participant recruitment and increased customer engagement through data visualizations. While grid-edge computing requires new hardware, cloud infrastructure more readily extends the computation and data storage capabilities of existing internet-connected devices, allowing more complex algorithms to be run offsite. Inverter-based CERs usually include measurement and communication capabilities, enabling the power consumption/production of these devices to be measured and communicated offsite. The additional information from these devices, grid-edge computing, and energy disaggregation~\cite{armel_disaggregation_2013,nationalGrid2022} allow richer measurement streams, which can be used to improve or augment DR capabilities.

Limitations of these technologies include cost, security, and reliability. As noted above, grid-edge computing requires new hardware, which can be expensive, and robust data communication networks~\cite{nationalGrid2022}. Limitations of cloud computing include cybersecurity concerns and costs. Additional sensing capabilities and energy disaggregation generally require new hardware and data communication capabilities. The (un)reliability of energy disaggregation estimates may also limit their usefulness. 

\subsubsection{Communication}
Opportunities to enhance DR adoption and scale include 1) increasing data throughput by improving communication network speeds, 2) seamlessly interconnecting systems through interoperability, and 3) securing communication networks through cybersecurity. Communication network speeds dictate the volume of information that can be passed between internet-connected hardware components. Interoperability is ``the ability to exchange actionable information between two or more systems and across and within organizational boundaries,"~\cite{interoperability_2016} and it defines the ease with which CERs can be integrated into a DR system. Cybersecurity refers to the ability to secure data flows, databases, and hardware against external threats and abuse. 

The deployment of 5G wireless communication technology increases data throughput by orders of magnitude~\cite{wireless_strategy_2022} creating opportunities for communicating more information and for transmitting information at higher frequencies. This enables 1) a better understanding of how the system is operating, 2) control on faster timescales, and 3) provision of more effective DR services. Faster communication speeds also support scaling DR, as more data must be transmitted in a given time frame as more CERs participate in a DR system. Creating seamless interoperability of CERs, AMI, supervisory control and data acquisition (SCADA) systems, cloud infrastructure, etc.\ reduces system integration costs and can allow a more diverse set of CERs (and CER manufacturers) to participate in DR. Securing communication systems reduces concerns that attacks on CERs might negatively impact power systems.

Concerns over communication delays, the performance of interoperability efforts, and the vast number of CERs that must have secure communication channels may limit the adoption of DR. Specifically, the latency in communicating between existing hardware and the cloud, along with security and cost considerations, may limit its application~\cite{cecRoadmap2021}. Specialized CER communication protocols exist to enhance interoperability; however, these protocols may be less reliable than other protocols (e.g., SCADA protocols) resulting in poor performance~\cite{cecRoadmap2021}. In addition, protocols can be difficult to use due to their complexity, reducing interoperability rather than enhancing it. Existing cybersecurity standards may need to be updated and refined for DR~\cite{wireless_strategy_2022}. Cybersecurity must mitigate the impacts of CER attacks on the broader power system~\cite{usCybersecurity}; otherwise it may be necessary to limit the scale of DR. 

Effective standards can help achieve these goals, and poor standards can hinder DR adoption and services. The development of standard API interfaces for CERs, standard and lightweight communication protocols for interacting within DR systems, and standards for cybersecurity would allow DR to operate with less overhead and with more security, lowering barriers to DR adoption and scale. Currently, most manufacturers have their own proprietary communication protocols and/or APIs, requiring DR operators to develop CER integration strategies for each device/manufacturer~\cite{cecRoadmap2021}. Standardization of protocols and security measures would enhance interoperability, reducing costs and time-to-deployment. It would also mitigate concerns about the vulnerability of these systems~\cite{usCybersecurity}. 

\subsubsection{Modeling and Forecasting}
Modeling and forecasting are necessary to establish baselines, i.e., estimates of what CERs would have consumed/produced under normal operation, without the influence of DR. In some cases, baselines are used to determine the performance of or payment for DR. Accurate baseline estimation remains a fundamental challenge~\cite{mathieu_quantifying_2011}. Some DR programs require participants to declare their own baseline in advance of events~\cite{chao2011demand}. This requires the participant, rather than the DR operator, to model and forecast the baseline. Some DR architectures do not require baselines, e.g., P2P trading and uni-directional pricing. 

Machine learning algorithms are improving the accuracy with which the behavior of CERs, both their baselines and responses, can be forecast.  Online learning algorithms can also be used to disaggregate network load from different types of CERs in real-time, leveraging typical SCADA and building-level data~\cite{ledvaDisagg}. Real-time estimates of the load from different types of CERs can be used to develop real-time estimates of CER flexibility. However, the reliability, data needs, and complexity of these approaches can be a barrier to adoption. 

While advanced modeling, forecasting, and learning approaches are being developed, CER responses are often seen as unreliable by grid operators, and the resources may not be fully utilized as a result~\cite{cecRoadmap2021}. In addition, many advanced algorithms, such as deep learning, rely on extensive amounts of data that might not be available, making the approaches impractical. As these algorithms become more complex, more specialized knowledge is required to implement/operate them, which may limit the personnel capable of working with them. In addition, models need to be explainable to DR operators and regulators, so that the impacts of different inputs to the models are properly understood. 

\subsubsection{CER Orchestration}
Enablers of CER orchestration include ensuring consumer comfort and increased automation. Specifically, most modern DR orchestration algorithms focus on minimizing disruption to the end-user, e.g., control of air conditioners that does not allow the indoor temperature to deviate outside of the normal operational range while allowing flexibility in when the device consumes power. Automation at the building level is becoming more common, allowing more effective and targeted responses to DR signals~\cite{buckberry2020smart}.  
However, barriers to the adoption include CER orchestration complexity and difficulty justifying the costs of enabling CER orchestration~\cite{cecRoadmap2021}. Specifically, increasingly complex DR applications and algorithms may be difficult for DR operators to implement or regulators to oversee. The cost of establishing control of CERs and verifying results may be prohibitive compared to the availability of other resources. 

Another key challenge is the impact of DR on distribution network constraints. As DR programs scale, a lack of awareness and consideration for the state of network may cause a decrease in system reliability, e.g., due to voltage violations or component overloading, and cause both DR operators and CER owners to distrust DR. Therefore, researchers have proposed ``distribution network-aware" CER orchestration approaches~\cite{buckberry2020smart, arpaeNODES}. However, these usually require integration with SCADA systems and/or other systems at the distribution level. A possible solution is to cast the problem as an optimal power flow problem and solve it in a centralized or distributed fashion; the latter approach was recently trialed on Bruny Island in Australia~\cite{Bruny_DOPF_2021}. However, this requires full CER and network observability, which is challenging for two reasons. First, the regulatory separation between DSOs and retailers (in Europe and Australia) and between utilities and aggregators (in North America) results in information asymmetry~\cite{ross_strategies_2021}, i.e., retailers and aggregators generally do not have access to network data, DSOs/utilities may not have access to CER data, and DSOs and aggregators may not have AMI data. 
Second, DSOs/utilities may not have detailed or accurate models of their low-voltage networks.

\subsection{Regulatory, Policy, and Market Design Considerations} \label{sec:MarketBarriers}
We next highlight barriers and enablers in today's regulations, policies, and market designs. In many jurisdictions, regulatory frameworks, such as FERC Order 2222 (US) and the Electricity Directive (EU), have recently rolled out to begin to address some of the issues listed below. 

\subsubsection{Lack of proper legal/regulatory frameworks} 
Enabling DR often requires the introduction of new actors, such as aggregators, as well as new relationships between new and existing actors. Legally defining these new entities, their rights (e.g., to participate in electricity markets) and responsibilities (e.g., in case of conflict with a consumer) is crucial to provide a basis for their activities. However, in the EU, four member states do not have a legal definition of an ``aggregation"~\cite{ACER2023}.  The absence of these legal frameworks is an entry barrier for new market participants, which in turn may limit competitive pressure and innovation~\cite{ACER2023}.  Similarly, with data playing a crucial role in many DR architectures, non-discriminatory and transparent rules and procedures on access, exchange, and use of (consumer) data must be put in place~\cite{ACER2023}. However, access to consumer data by aggregators is not yet possible in Spain, Italy, and Poland~\cite{ACER2023}. Legal and regulatory frameworks should allow all resources and actors to participate in electricity markets, provided they meet technical and financial requirements. However, aggregations and CERs are not allowed to offer balancing services in Denmark~\cite{ACER2023} and, prior to FERC Order 2222, FERC-regulated US states could choose to ban aggregators. Now DERs are or will be allowed to participate in FERC-regulated US electricity markets, but some regions (e.g., MISO) may wait until 2029~\cite{PNNL2022}. 

\subsubsection{Challenges with remuneration}
Measuring CER response to DR signals, required by most DR remuneration schemes, requires direct CER power measurements and/or AMI. However, the roll-out of AMI is incomplete, e.g., in 2022, only 13 out of 27 EU member states had AMI deployment levels of over 80\%~\cite{ACER2023}. Furthermore, AMI does not always have the necessary features to fully harness the potential of CERs. Another renumeration challenge is that, for some DR services, markets/incentives are missing, e.g., markets for local network services are non-existent (except for some early trials~\cite{ARENA_DEIP_2022}). Cost-based congestion management, favored by many transmission system operators in Europe~\cite{ACER2023}, typically elicits services from conventional power plants, even though CERs may provide these services more cost-effectively.

Current retail rate structures typically mute the benefits of DR. Adoption of real-time pricing, which would better signal the need for and value of DR to consumers, is slow. In part, this may be driven by concerns about excessive (variations in) electricity bills as seen in the 2021 Texas Energy Crisis~\cite{griddy}. A notable exception is Norway, where over 90\% of consumer have opted for dynamic electricity price contracts. Furthermore, the energy component in retail rates is often low (in 2022, it represented, on average, less than 50\% of the final electricity bill for a household in 10 EU member states~\cite{ACER2023}), muting the benefit of shifting demand. Innovative distribution and transmission network tariffs, which could, e.g., signal the value of local network services, are lacking in many parts of the world. For example, only Spain, Norway, and Austria use time-differentiated network tariffs for all consumers~\cite{ACER2023}. Furthermore, regulatory interventions and policies, though well-intentioned, further distort the incentives for DR. Examples include capping retail rates (e.g., during the 2022 energy crisis in Europe), social tariffs for vulnerable consumers, subsidy schemes, and net-metering policies. 

\subsubsection{Requirements that limit market access}
CER access to electricity markets is often restricted or impeded by regulatory and participation requirements, such as minimum bid sizes and long market lead times. For example, for US ISO-run day-ahead energy markets, CER aggregations are typically represented through convex offer curves or simple price-quantity pairs~\cite{PNNL2022,ESIG2022}. More complex offer structures would allow more accurate representation of constraints and costs~\cite{PNNL2022,ESIG2022}. In the European day-ahead market, CER aggregators may use complex bid and order structures to capture the limitations and costs of the underlying assets~\cite{Euphemia}. However, excessive administrative and financial requirements (e.g., collateral) hamper CER market participation~\cite{ACER2023}. 

Several design features of the European operating reserve markets, which are cleared separately from the day-ahead energy market, limit the uptake of DR. Examples include non-market-based reserve capacity, restrictions in the pre-qualification phase, large minimum bid or capacity sizes, long procurement lead times, and long balancing capacity contracts~\cite{ACER2023}. As a result, DR provided less than 5\% of the balancing capacity across all reserve categories in all EU member states except France in 2022. In the US, after the implementation of FERC Order 2222, DERs will be able to provide reserve capacity though participation requirements may still make that challenging for some CERs~\cite{PNNL2022}.

Capacity markets may impose CER participation restrictions   through product/contract design or in the capacity allocation process. For example, DR may not be eligible to participate in a capacity market due to requirements on the connection voltage level or minimum asset size, such as in Germany's strategic reserve~\cite{ACER2023}. Multi-year contracts with year-round availability requirements, e.g., in the Belgian capacity market, are more challenging to fulfill by CERs~\cite{ACER2023}. The allocation process, typically an auction, may feature long lead times, minimum bid sizes, low capacity, or DR derating factors. As a result, the share of DR in contracted volumes has typically been below 10\% in the European capacity markets~\cite{ACER2023}.

\subsection{Social Considerations} \label{sec:SocialBarriers}

The grid is a complex subject, and most of the general public has a highly simplified view of power systems. Their only personal interaction with electricity has, ironically, been in its absence: when an electric appliance breaks, a phone battery runs out, or a home experiences a blackout. That is, people think about electricity during the 0.8\% of the year it is unavailable~\cite{NERC_SOR2022}. Thus, it is no wonder that DR programs receive very little interest and struggle to achieve scale. Many DR operators provide minimal DR/grid education to consumers, still rely on crude incentive mechanisms (e.g., \$25 to sign up) that are poorly communicated (if at all), mischaracterize the dynamic nature of DR as responsiveness only during discrete and rare ``events,'' and do not equitably share the benefits of participation, resulting in recruitment and retention challenges. Moreover, DR incentives are generally very small, making participation ``not worth thinking about.''

In addition, many existing opt-in DR programs indirectly favor participants who can afford new (more expensive) CERs and skew participation to the upper socioeconomic, more senior, and more highly-educated segments of the population~\cite{MortonIowa2021,THORSNES2012552}. Given that the direct benefits from participating are financial and shared only with participants, DR programs could shift costs from more well-off (who can afford fuel-switching/upgrading appliances) to less well-off consumers (who cannot afford upgrades), raising concerns around energy equity and justice~\cite{cong2022unveiling, mathieuEJ:2023}.  Insufficient communication between DR operators and consumers also causes distrust in terms of a perceived lack of data security and privacy, and displeasure/discomfort from participating~\cite{MILCHRAM20181244,OCONNELL2014686}.

Recent public policies are trying to shift the status quo by more actively including diverse communities in conversations about the power system~\cite{BlocPower_Baird2023} and by promoting decentralization and democratization of energy systems, such as with \textit{energyshed} design~\cite{Thomas2021EShed,hamilton2023towards}, as means to achieve decarbonization and electrification objectives. As a result, some DR operators are now placing consumers at the center of DR programs~\cite{GMP_Vermont2021}. Over the last decade, this has led to highly-subsidized (or free) smart devices being made available in energy-as-a-service subscription programs (e.g., rent 10~kW of batteries for \$55/month~\cite{gmp_green_2020}), automated enrollment schemes, over-the-air updates, cohesive messaging about global (CO$_2$) and local (energy cost) impacts, and advanced CER orchestration that accounts for comfort/quality of service. Some DR technologies have advanced so rapidly that the DR operators themselves are struggling to keep up, which causes a growing distrust among CER manufacturers, who are concerned with consumers having a poor DR experience. In sum, more conversations are necessary between device manufacturers, DR operators, regulators, and the general public to overcome social barriers and scale DR efforts. Our new DR definition provides a common, clear starting point. 

\subsection{A Demand Response Research Agenda}

Based on the prior discussions, we have identified  technical, regulatory/economic, and social challenges and needs around DR, requiring more research, development, and deployment.

\subsubsection{Technical}
{\bf Robust communication networks} are essential for communication between CERs, the cloud, and grid operators/utilities/retailers. This requires advancements in interoperability, cybersecurity, cloud communication, grid-edge computing, and network reliability. Moving towards {\bf ``plug and play" capabilities} can reduce interoperability barriers. Regulators can enforce {\bf standardization} to reduce costs associated with device APIs. {\bf Grid-edge computing} can reduce reliance on broadband systems but incurs hardware, deployment, and maintenance costs. Reducing these costs can accelerate DR adoption. {\bf 
Improved models and forecasts} of CER behaviors enhance DR utilization. Transparency in DR performance reduces expectations around response uncertainty. Clear model and forecast output specifications can enhance DR adoption. CER orchestration is limited by {\bf lack of integration with SCADA} and high device manufacturer {\bf API limitations and fees}.

\subsubsection{Regulatory/Economic}
There is a need for {\bf methods to enable proper remuneration}, reflecting the true value of DR. This includes {\bf advancing rate and market designs}, including promoting real-time pricing and advanced network tariffs, removing distortive regulatory interventions, creating novel markets to unlock untapped DR value (e.g., local network services), and removing restrictive market access requirements in existing markets. The latter requires {\bf defining roles and responsibilities}, including removing regulatory and legal barriers to allow CERs to offer DR, directly or through aggregators, in all electricity markets. Utilities should be able to {\bf rate-base investments in DR-enabling infrastructure}, e.g., sensing/communication assets. 
  
\subsubsection{Social} Public understanding of the electricity system lags behind that of many other technologies. Therefore, {\bf education and outreach} are essential. Public perception of utility companies is often poor, with inherent distrust of new technologies and programs, requiring {\bf trust-building} and/or involving other trusted entities in DR implementations.  DR schemes should {\bf ensure equitable participation and remuneration} among participants, and equitable {\bf sharing of value} among all consumers. Additionally, {\bf DR participation needs to be easier}, with simple sign-up processes, automated control, and clear and compelling remuneration schemes.

\section{Conclusion}

This paper proposed a new definition of DR that encapsulates the demand side's full capabilities and value in responding to power system needs. We demonstrated how existing definitions are largely insufficient, especially in the era of prolific DERs, and how DR-adjacent concepts clutter the landscape. Our definition aims to be as broad as possible while ruling out actions that are not DR, precise and concise in language, and, eventually, broadly accepted by the power systems community. Through a discussion of DR services and value streams, mechanism designs, and enabling technologies, we demonstrated the value of our new definition of DR. We also discussed DR barriers and enablers,  which allowed us to outline a DR research agenda.


\bibliographystyle{IEEEtran}
\bibliography{DR_DER_era_bibliography}

\end{document}